\documentclass[twocolumn,times,tighten]{aastex63}
\usepackage{amssymb,amsmath,mathrsfs,graphicx}

\newcommand{\tsd}{\tau_{\text{sd}}}
\newcommand{\La}{L_{\alpha}}

\begin{document}

\title{Evidence for magnetar precession in X-ray afterglows of gamma-ray bursts}

\author{A. G. Suvorov}\thanks{arthur.suvorov@tat.uni-tuebingen.de}
\author{K. D. Kokkotas}
\affil{Theoretical Astrophysics, IAAT, University of T{\"u}bingen, Germany}

\begin{abstract}
\noindent{Many gamma-ray bursts are followed by periods of extended emission. At least in some cases, the burst afterglow may be powered by a rapidly rotating, highly-magnetised neutron star, which spins down due to electromagnetic and gravitational wave emission. Such a remnant is likely to strongly precess in the early stages of its life, which would lead to modulations in the X-ray luminosity as the triaxiality of the system evolves over time. Using a radiation profile appropriate for a precessing, oblique rotator, we find that Swift-XRT data of a long (080602) and a short (090510) burst matches the model with significantly higher accuracy (mean-square residuals dropping by $\gtrsim 200 \%$ in the early stages of the extended emission) than for an orthogonal rotator. We interpret this as evidence for precession in newborn magnetars.}
\end{abstract}


\section{Introduction} 

A common feature of gamma-ray bursts (GRBs) is the emergence of a shallow decay phase following the prompt emission \citep{nou06}. The absence of spectral evolution during this `plateau' suggests that the afterglow is powered by continuous energy injections from a long-lived remnant \citep{zhang06}, born out of gravitational collapse or a compact object merger \citep{piran04,berg14}. Depending on the properties of the progenitor star(s), a remnant in the form of a rapidly rotating, highly-magnetised neutron star (`millisecond magnetar') is expected. In this scenario, charged particles near the remnant may be rapidly accelerated towards the surrounding ejecta by magnetic winds, eventually shocking the envelope and collimating a blast wave \citep{met08,dall11,bucc11}. The nascent neutron star then continues to fuel the system by stabilising itself via electromagnetic and gravitational wave emission \citep{fan13,kokk15,gao16,sarin20}, {which may contribute to the synchrotron emission from the expanding `fireball' as the GRB jet interacts with the surrounding interstellar medium} \citep{mez93,sarin19}. The afterglow luminosity gradually decays as the star decelerates over a spin-down time-scale $\tsd$. The magnetic field strength, equation of state, and rotational frequency can thus all be studied by comparing the theoretical radiation luminosities with the observed light curves \citep{rowl14,lasky16,stratta18}.

Interestingly, the X-ray afterglow curves for certain GRBs display oscillatory behaviour on time-scales shorter than $\tsd$ \citep{der99,farg03,marg08}. If these oscillations are genuine features of the energy profile of the remnant rather than of observational or astronomical systematics, {such as red noise at $\sim$ Hz frequencies due to the variability of the burst itself} \citep{gold17}, it is possible that even more information can be extracted about the nature of GRB remnants \citep{marg19}. 

The efficiency of both electromagnetic and gravitational radiation emitted by a newborn star depends on the details of its orientation and triaxiality \citep{xiao19}, most notably on the inclination angle $\alpha$ made between the rotation $\boldsymbol{\Omega}$ and magnetic $\boldsymbol{B}$ axes \citep{mus19}. Since post-formation convection is expected to erase any pre-existing correlation between these two vectors \citep{thom93}, the remnant is likely an oblique rotator, at least at birth \citep{melatos00,lander18}. In general, $\alpha$ evolves over time as $\boldsymbol{\Omega}$ and $\boldsymbol{B}$ evolve, and oscillations or `wobbles' in $\alpha$ are a feature often seen in magnetohydrodynamic simulations \citep{gog15,lai15,arz15}. If the star is rapidly rotating and highly-elliptical, early-time precession may then modulate the spin-down luminosity to some extent on time-scales less than $\tsd$. It is the purpose of this letter to demonstrate that fitting a precessing magnetar model to Swift-XRT data \citep{evans09} results in significantly lower residuals than for an orthogonal rotator, at least in some cases. We model the emission profiles for remnants in GRB 080602 (a long GRB) and GRB 090510 (a short GRB), as they both display the `plateau' phase expected of a magnetar central engine with strong dipole field \citep{rowl13,stratta18}.
 
A concordance cosmology with \cite{planck18} parameters $H_{0} = 67.4 \text{ km s}^{-1} \text{ Mpc}^{-1}$, $\Omega_{M} = 0.32$, and $\Omega_{\Lambda} = 0.68$ is adopted throughout to translate between measured fluxes and (bolometric) source luminosities. In Section 2 we briefly review the magnetar central engine model for extended emission in GRBs, and compare the well-known formula for electromagnetic spin-down for an orthogonal rotator with a generalised expression appropriate for a precessing, oblique rotator. Light curve fits for GRBs 080602 (Sec. 3.1) and 090510 (Sec. 3.2) are then presented for both models, with some discussion given in Section 4.

\section{Millisecond magnetar engines for GRBs}

The energy reservoir of a newborn neutron star consists primarily of its rotational kinetic energy\footnote{The convention $Z_{x} = 10^{-x} Z$ in CGS units, with the exception of mass measured in solar masses, is adopted throughout.},
\begin{equation} \label{eq:energy}
E_{\text{rot}} = \frac {1} {2} I \Omega^2 \sim 2 \times 10^{52} M_{1.4} R_{6}^{2} P_{-3}^{-2} \text{ erg},
\end{equation}
where $I \sim 2 M R^2/5$ is the moment of inertia for stellar mass $M$, radius $R$, and spin period $P = 2 \pi / \Omega$.  In the magnetar central engine model for GRBs, the X-ray afterglow following the prompt emission is powered by the injection of energy into the forward shock through the conversion of mechanical energy \eqref{eq:energy} into radiation energy with luminosity $L$,
\begin{equation} \label{eq:spin-downrot}
-\dot{E}_{\text{rot}} = - I \Omega \dot{\Omega} = \eta L,
\end{equation}
{where $\eta \leq 1$ is an efficiency parameter accounting for imperfect conversion [see e.g. \cite{xiao19}]}. If the newborn star is an oblique rotator with millisecond period and strong ($\gtrsim 10^{15} \text{ G}$) magnetic field, the dominant term within $L$ may be sourced by electromagnetic braking. In general, neutrino outflow from Urca cooling \citep{thom04}, {fall-back accretion \citep{mel14}}, or gravitational radiation may also be important \citep{cors09} as substantial, time-varying quadrupole moments can be induced by magnetic deformations \citep{mast15} or quasi-normal oscillations \citep{kru19}. We will ignore these effects here for simplicity (though see Sec. 3).

\subsection{Electromagnetic dipole radiation}

In the dipole approximation, the electromagnetic spin-down luminosity associated to a neutron star with polar field strength $B_{p}$ is given by
\begin{equation} \label{eq:spin-down}
L_{\text{EM}} =  \frac {B_{p}^2 R^{6} \Omega^4} {6 c^3} \lambda(\alpha),
\end{equation}
where $\lambda$ depends on the neutron star orientation through the inclination angle $\alpha$ and the magnetospheric physics. In vacuum one has $\lambda(\alpha) = \sin^{2}\alpha$, though the neutron star is unlikely to be perfectly isolated in reality, as charges accumulate in the magnetosphere via induction-generated electric fields near the stellar surface \citep{gj69}, especially if there are active magnetic winds; numerical simulations of charge-filled magnetospheres suggest instead that $\lambda(\alpha) \approx 1 + \sin^2 \alpha$ in reality \citep{spit06,kara09,phil15}. Here, we adopt a hybrid model with $\lambda = 1 + \delta \sin^2 \alpha$, where the parameter $|\delta| \leq 1$ quantifies our ignorance of the magnetospheric physics \citep{arz15}.

In general, the energy-balance equation \eqref{eq:spin-downrot} can be solved using expression \eqref{eq:spin-down} to determine the angular velocity as a function of time, which then determines the X-ray luminosity $L(t)$. In the case $\delta = 0$ (equivalently, for an orthogonal rotator in vacuum with $\alpha = \pi/2$), $L$ takes the well-known form
\begin{equation} \label{eq:spiinlum}
L_{\perp} = \frac {B_{p}^2 R^6 \Omega_{0}^4} {6 c^3} \left( 1 + \frac {t} {\tsd} \right)^{-2},
\end{equation}
where $\tsd \sim 2.3 \times 10^{3} \text{ s} \times \left( M_{1.4} B_{p,15}^{-2} P_{0,-3}^{2} R_{6}^{-4} \right) $ is the characteristic spin-down time for birth period $P_{0} = 2 \pi / \Omega_{0}$. GRB afterglows that are characterised by a roughly constant `plateau' phase ($t \ll \tsd$) followed by a $t^{-2}$ falloff $(t \gg \tsd)$ are then well-described by \eqref{eq:spiinlum}. However, as initially argued by \cite{thom93} [see also \cite{melatos00,lander18}], turbulent convection in the post-merger (short GRBs) or post-collapse (long GRBs) remnant breaks any pre-existing correlation between the Euler angles of the system, and thus it is unlikely that $\boldsymbol{\Omega} \cdot \boldsymbol{B} \approx 0$ everywhere at birth, regardless of environmental details.

\subsection{Precession}

At times $t$ less than the spin-down time $\tsd$, the evolution of the inclination angle $\alpha$ is mainly driven by precession \citep{gold70,lai15},
\begin{equation} \label{eq:alphaeqn}
\dot{\alpha} \approx k \Omega_{p}  \csc \alpha \sin(\Omega_{p} \times t ),
\end{equation}
where $k$ is an order-unity factor which is related to the other Euler angles defining the triaxial neutron star and $\Omega_{p} = 2 \pi / P_{p}$ is the precession velocity, related to the rotational velocity through $\Omega_{p} \approx \epsilon \Omega$ for oblateness $\epsilon$ [see e.g. \cite{gog15}], which in turn is related to the mass quadrupole moment of the newborn magnetar [see e.g. \cite{jar98}]. 

Assuming that the oblateness remains constant until several spin-down times have elapsed, we obtain an approximate solution to the coupled system \eqref{eq:spin-down} and \eqref{eq:alphaeqn}, which implies that the precession-modified electromagnetic spin-down luminosity reads 
\begin{equation} \label{eq:genspinlum}
\begin{aligned}
\hspace{-0.1cm}\La \approx&  \frac {B_{p}^2 R^6 \Omega_{0}^4} {6 c^3}  \left\{ 1 + \delta - \delta \left[ \alpha_{0} + k \cos \left( \Omega_{p} \times t \right) \right]^{2} \right\} \\
&\times \Big\{ 1 + \frac {t \left[ 1 + \delta \left( 1 - \alpha_{0}^2 - \tfrac {1} {2} k^2 \right) \right]} {\tsd} \\
&- \frac {k \delta \left[ 2 \alpha_{0} + \tfrac{1}{2} k \cos \left( \Omega_{p} \times t \right) \right] \sin \left( \Omega_{p} \times t \right)} {\tsd \Omega_{p}} \Big\}^{-2},
\end{aligned}
\end{equation}
where $\alpha_{0}$ is related to the inclination angle at birth, and
\begin{equation} \label{eq:precessionfreq}
\Omega_{p}(t) \approx \epsilon \Omega_{0} \left( 1 + \frac {t} {\tsd} \right)^{-1/2}.
\end{equation}
Expression \eqref{eq:genspinlum} generalises that of \eqref{eq:spiinlum} by including magnetospheric physics ($\delta$), the Euler angles at birth ($k, \alpha_{0}$), and precession ($\Omega_{p}$), which are not entirely independent. In any case, setting $\delta = 0$ returns the orthogonal rotator solution \eqref{eq:spiinlum}. The physical parameters of the nascent neutron star can be then be inferred by fitting afterglow data to the spin-down luminosity \eqref{eq:genspinlum}.


\section{Light curve fitting}

We employ simple error-weighted Monte Carlo simulations to minimise the square residuals when fitting a spin-down luminosity to afterglow data from the Swift-XRT catalogue \citep{evans09}. To this end, we consider two models here: the orthogonal rotator, for which the spin-down luminosity is given by the well-known formula \eqref{eq:spiinlum}, and the oblique, precessing rotator, where the luminosity is instead given by expression \eqref{eq:genspinlum}.

It is important to note that the latter model described above has more free parameters than the former. This implies that we should always obtain a better fit, regardless of whether the additional parameters are physically important or not. To provide evidence that the additional parameters are not spurious but rather represent meaningful physics, we can calculate the associated Akaike information criterion (AIC) of each fit, where
\begin{equation} \label{eq:aicn}
\text{AIC} \propto  m - \underset{m}{\text{max }} \hat{\ell} ,
\end{equation}
for $m$ parameters to be estimated, where $\hat{\ell}$ is the log-likelihood function of the model. In general, given a set of candidate models for some data, the one with the minimum AIC value is preferred \citep{akaike}. In both cases discussed below, we find that the AIC number for the precession model is substantially smaller than for the orthogonal rotator. {A summary of fitted parameters is given in Table \ref{tab:fits} below.}

\subsection{A long GRB: 080602}

GRB 080602 is a long GRB at redshift $z = 1.82$ \citep{kruh15}, with prompt burst duration $T_{90} = 74 \pm 7 \text{ s}$ in the $15-350$ keV band {and photon index $\Gamma \approx 1.43$}. In general, there is a delay of several tens of seconds between the prompt emission and the afterglow, which may be attributed to the activation time of the magnetic winds: it takes $\gtrsim 10$ seconds for the (proto-)magnetar to cool enough for the neutrino-loaded magnetic wind to become ultra-relativistic \citep{met08}. Following the onset of extended emission, a steep decay in the flux is observed until $t \sim 150 \text{ s}$ after the Swift Burst Alert Telescope (BAT) trigger, interpretable as the fading of the impulsive energy provided by the adiabatic fireball \citep{cors09,bucc11}. Direct fits to the observed X-ray flux $F$, which is related to the luminosity through $L(t) = 4 \pi D_{L}^2 F(t) K(z)$ for luminosity distance $D_{L}$ and {cosmological $k$-correction factor $K(z) = (1+z)^{\Gamma-2}$} \citep{bloom01,mus19}, are shown in Fig. 1. A zoom-in of the first $150$ seconds post-fireball are shown in Fig. 2 for improved visibility. {In this section, we assume canonical values of $R = 12 \text{ km}$ and $M = 1.4 M_{\odot}$.}

\begin{figure}[h]
\includegraphics[width=0.473\textwidth]{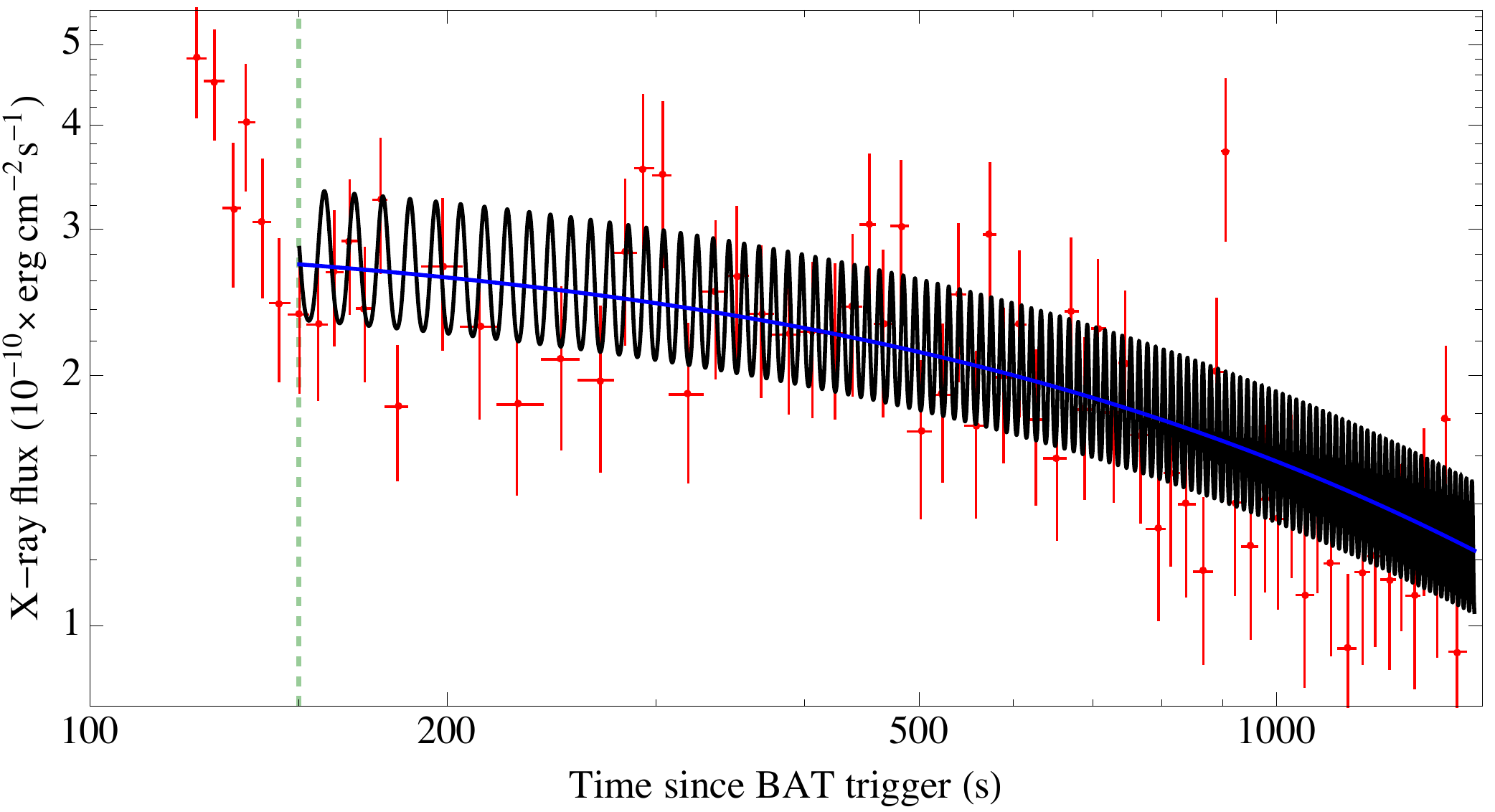} \label{fig:080602}
\caption{Light curve fits for GRB 080602 (data shown in red; Evans et al. 2009) via electromagnetic spin-down from a newborn magnetar, where the impulsive fireball energy ends $\sim 150$ seconds (dashed, green line) after the initial detection by the Swift BAT. Over-plotted are the best fits of the data to an emission profile for an orthogonal rotator in vacuum (blue curve) and for a star with a magnetic inclination angle evolving due to precession (black curve). The oscillations become increasingly damped as the precession frequency decreases.}
\end{figure}

\begin{figure}[h]
\includegraphics[width=0.473\textwidth]{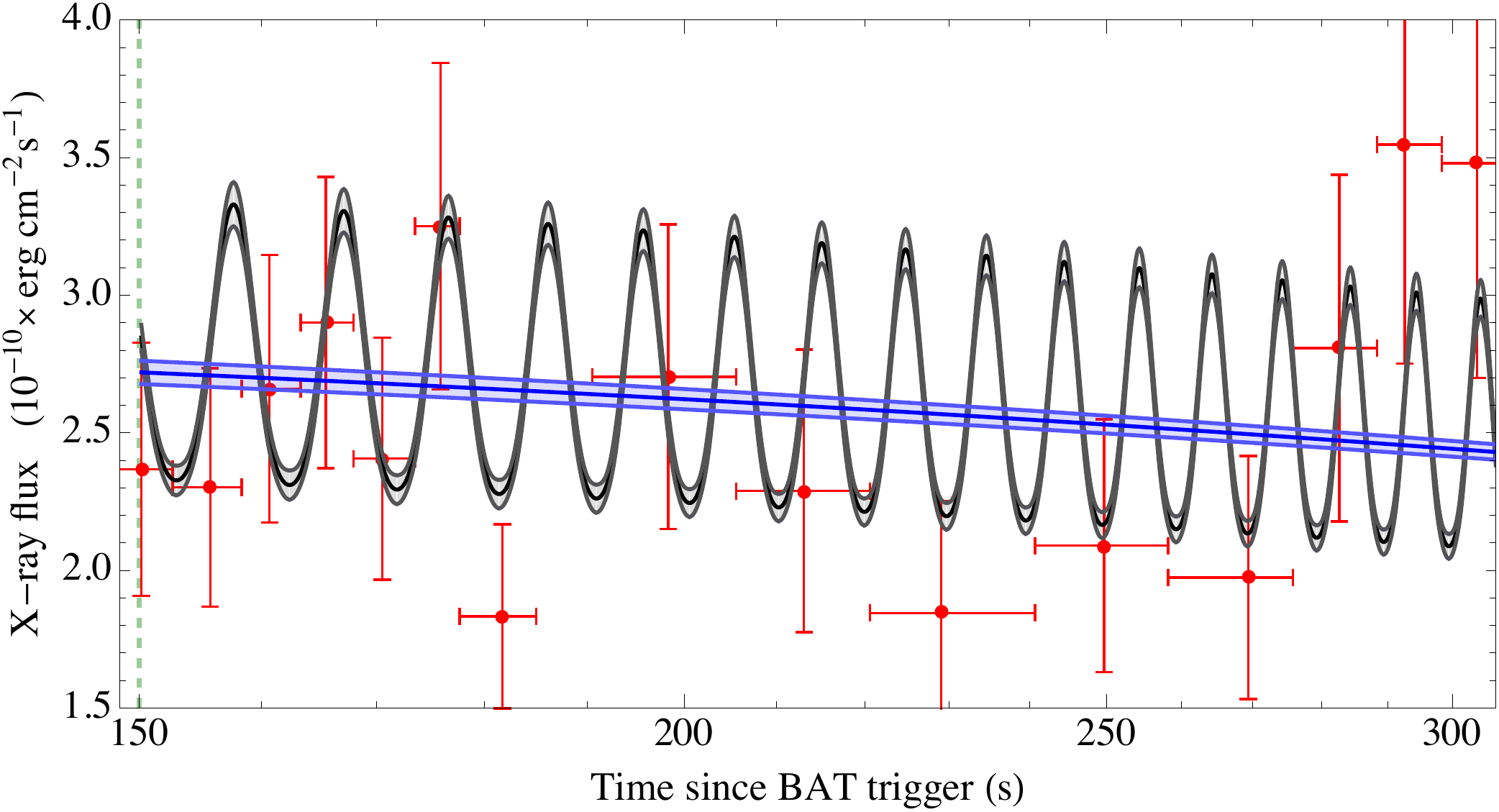} \label{fig:080602zoom}
\caption{A zoom-in of the first $150$ seconds post-fireball for the fits shown in Fig. 1. The shaded regions surrounding the blue and black curves indicate the respective $90\%$ confidence intervals for the fits.}
\end{figure}

For the precessing model, the best fit numbers are found as: $B_{p}/\sqrt{\eta_{0.5}} = 1.02 \times 10^{15} \text{ G}$, $P_{0}/\sqrt{\eta_{0.5}}  = 1.25 \text{ ms}$, $\delta = -0.46$, $\alpha_{0} = -0.47$, $k = 0.29$, and $\epsilon/\sqrt{\eta_{0.5}} = 1.45 \times 10^{-4}$. {Note that adjusting the efficiency $\eta$ scales the polar field strength and spin period accordingly.} In general, an ellipticity of the order $\epsilon \sim 6 \times 10^{-6} B_{p,15}^2 R_{6}^{4} M_{1.4}^{-2}$ is expected from magnetic deformations alone \citep{mast15}. As such, if the internal field is $\lesssim4$ times stronger than the polar $B_{p}$ value, as suggested by galactic magnetar observations \citep{kaspi17}, or if there are active quasi-normal oscillations \citep{kokk15,lasky16,kru19}, the model pulls out the anticipated ellipticity quite organically.

Defining the (normalised) averaged mean-square residuals $\sigma$ over $N$ data points as
\begin{equation}
\sigma=  \frac {1} {N} \sum^{N}_{i=1} \left[ F_{-10}(t_{i}) - F_{\text{data,-10}}(t_{i}) \right]^{2},
\end{equation}
we obtain $\sigma_{\alpha} = 0.17 \pm 0.07$ over the entire data set for this model, while for the first 150 seconds (though with the same fit) we find $\sigma_{\alpha} = 0.15 \pm 0.13$.

For the orthogonal rotator we find $B_{p}/\sqrt{\eta_{0.5}}  = 7.91 \times 10^{14} \text{ G}$ and $P_{0}/\sqrt{\eta_{0.5}} = 1.21 \text{ ms}$. In this case, $\sigma_{\perp} = 0.24 \pm 0.08$ over the entire data set, while for the first $150$ seconds we have $\sigma_{\perp} = 0.33 \pm 0.19$. As such, the fit is improved by $\approx 40\%$ over the entire data set when using $\La$ over $L_{\perp}$, while in the early stages of emission, where precession is most important, the fit is improved by $\approx 220 \%$. These improvements, together with the AIC numbers {(see Table \ref{tab:fits})}, suggest that a precessing model is favoured over the standard spin-down model.

\subsection{A short GRB: 090510}

GRB 090510 is a short GRB at redshift $z= 0.9$ \citep{ack10}, with prompt emission duration $T_{90} = 0.3 \pm 0.1 \text{ s}$ in the $15-350$ keV band {and photon index $\Gamma \approx 0.98$}. {As this object was likely born out of a merger event \citep{berg14}, we assume the star is more compact, and take values $R = 12 \text{ km}$ and $M = 2.0 M_{\odot}$.} For this GRB, no impulsive-fireball phase is explicitly evident in the Swift-XRT data, though a plateau is seen at early times, eventually decaying as $\sim t^{-2}$ \citep{evans09}. Direct fits to the observed X-ray flux $F$ are shown in Fig. 3, while a zoom-in of the first $80$ seconds of data are shown in Fig. 4. After a few $\tsd$ have elapsed, the `wobbling' settles down, and the oscillation amplitudes decrease.

\begin{figure}[h]
\includegraphics[width=0.473\textwidth]{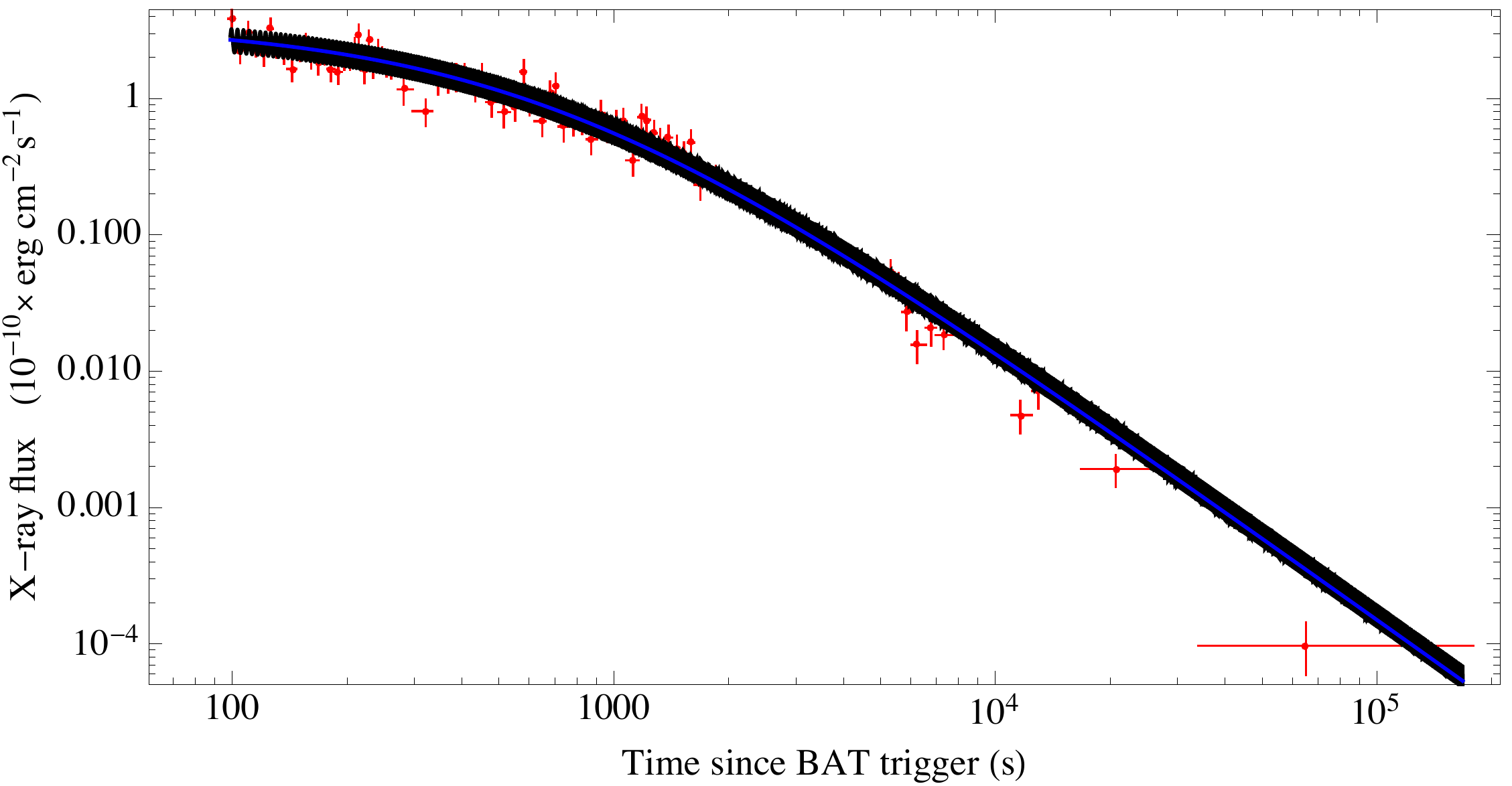} \label{fig:090510}
\caption{light curve fits for GRB 090510 (data shown in red; Evans et al. 2009) via electromagnetic spin-down from a newborn magnetar. Over-plotted are the best fits of the data to an emission profile for an orthogonal rotator in vacuum (blue curve) and for a star with a magnetic inclination angle evolving due to precession (black curve).}
\end{figure}

\begin{figure}[h]
\includegraphics[width=0.473\textwidth]{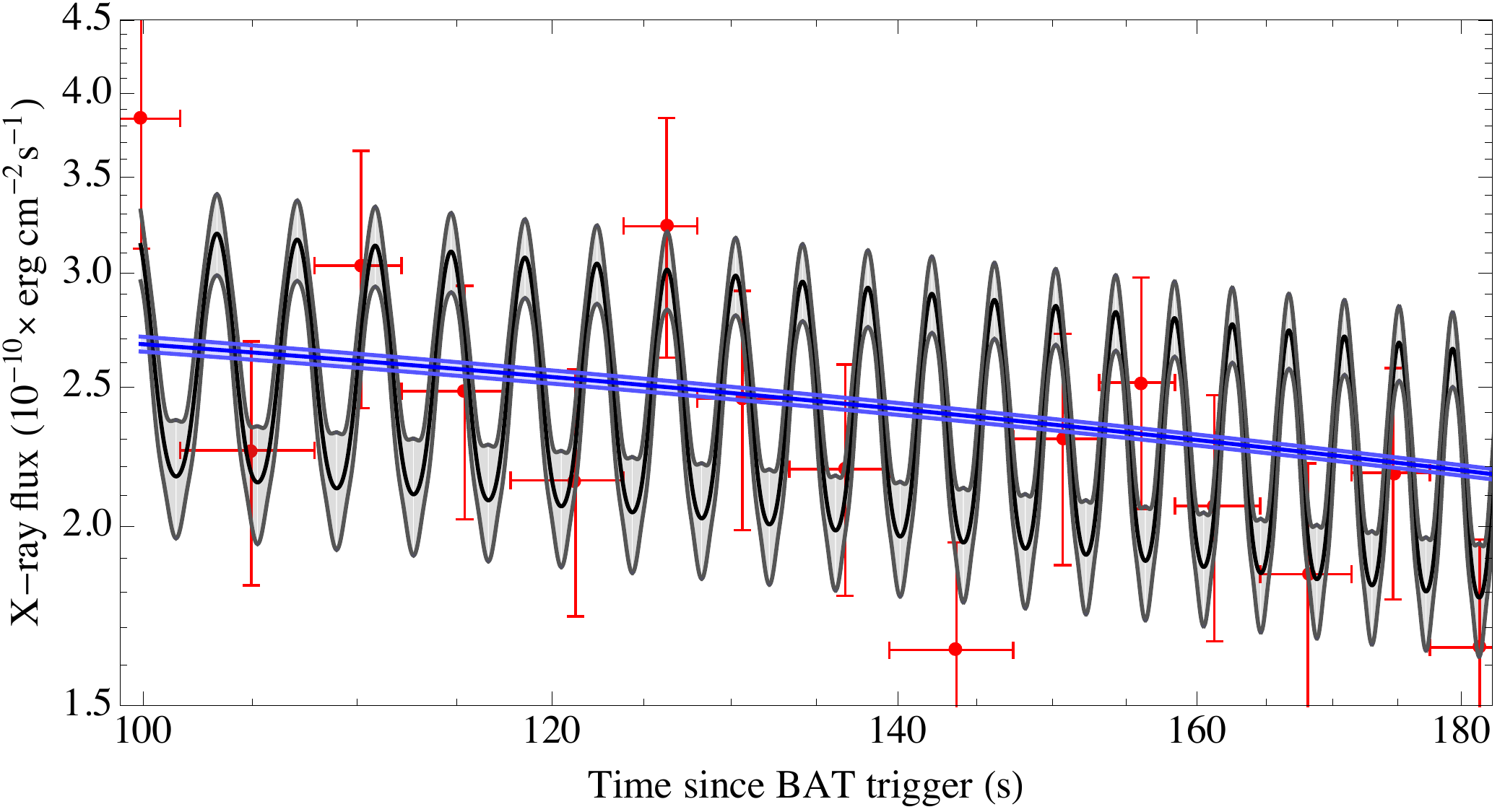} \label{fig:090510zoom}
\caption{A zoom-in of the first $80$ seconds of data for the fits in Fig. 3. The shaded regions surrounding the blue and black curves represent the respective $90\%$ confidence intervals for the fits.}
\end{figure}

For the precession model, the best fit parameters we find are: $B_{p}/\sqrt{\eta_{0.1}}  = 5.63 \times 10^{15} \text{ G}$, $P_{0}/\sqrt{\eta_{0.1}}  = 2.94 \text{ ms}$, $\delta = -0.51$, $\alpha_{0} = -0.49$, $k = 0.25$, and $\epsilon/\sqrt{\eta_{0.1}}= 9.65 \times 10^{-4}$. Again, an internal field of strength $B_{\text{int}} \lesssim 3 B_{p}$ would naturally induce an ellipticity of this order \citep{mast15}. The residuals read $\sigma_{\alpha} = 0.063 \pm 0.026$ over the entire data set, while for the first $80$ seconds we have $\sigma_{\alpha} = 0.11 \pm 0.09$. For the orthogonal rotator we find $B_{p}/\sqrt{\eta_{0.1}} = 3.22 \times 10^{15} \text{ G}$ and $P_{0}/\sqrt{\eta_{0.1}} = 2.48 \text{ ms}$ [which are similar to those obtained by \cite{rowl13}] while $\sigma_{\perp} = 0.11 \pm 0.04$ over the entire data set, and $\sigma_{\perp} = 0.24 \pm 0.14$ over the first $80$ seconds. Similar to the case of GRB 080602, the precession fits are $\approx 75\%$ better over the entire data set, though an improvement of $\approx 220\%$ is seen during the first $80$ seconds of data.

It is perhaps unsurprising that for a short GRB, possibly originating from a merger event, more extreme neutron star parameters are found than for the long GRB 080602; using numerical simulations, \cite{ross06} found that Kelvin-Helmholtz instabilities occurring at the shear layer between the progenitor stars can produce ultra-strong fields $(\lesssim 10^{17} \text{ G})$. Furthermore, the comparatively large value of the ellipticity $\epsilon$ is also unsurprising, since the remnant may be highly-deformed \citep{kokk15,gao16,sarin20}. Note that even for $\epsilon \sim 10^{-3}$, the gravitational-wave power is an order of magnitude smaller than the electromagnetic counterpart for $B_{p} \lesssim 10^{16} \text{ G}$ \citep{lasky16}.

\begin{table}
\caption{Properties of the fits obtained for GRBs 080602 (Sec. 3.1) and 090510 (Sec. 3.2), where we have assumed slightly different values for the efficiency $\eta$; see text for details. The AIC numbers are computed from \eqref{eq:aicn} with proportionality factor $2$ \citep{akaike}.}
\setlength\tabcolsep{2.5pt}
\hspace{-2cm}\begin{tabular}{c|c|c|c|c|c|c|c}
  \hline
  \hline
080602 & $B_{p}$ ($10^{15} \text{ G}$) & $P_{0}$ (ms) & $\delta$ & $\alpha_{0}$ & $k$ & $\epsilon$ ($10^{-4}$) & AIC \\
\hline
$L_{\alpha}$ & 1.02 & 1.25 & -0.46 & -0.47 & 0.29 & 1.45 & 17.1 \\
$L_{\perp}$ & 0.79 & 1.21 & - & - & - & - & 51.2 \\
\hline
090510 & & & & & & \\
\hline
$L_{\alpha}$ & 5.63 & 2.94 & -0.51 & -0.49 & 0.25 & 9.65 & 84.1 \\
$L_{\perp}$ & 3.22 & 2.48 & - & - & - & - & 96.5 \\
\hline
\hline
\end{tabular}
\label{tab:fits}
\end{table}

\section{Summary}

In this article, we generalise the standard electromagnetic spin-down luminosity \eqref{eq:spiinlum} for an orthogonal rotator to include the effects of precession, which naturally leads to fluctuations in the radiation luminosity $L$ during the first $\lesssim$ hours of a millisecond neutron star's life as the magnetic inclination angle $\alpha$ `wobbles' \citep{melatos00,gog15,lai15}. The wobbling may then manifest as damped oscillations in the X-ray fluxes seen from GRB afterglows powered by energy-injection from newborn millisecond magnetars (see Figs. 2 and 4). We find that these oscillations are consistent with observed short-time variabilities in GRBs 080602 and 090510, assuming that the variability is due to the nature of the source and not of instrumental or astronomical systematics, {such as from red noise due to the variability of the burst itself at $\sim$ Hz frequencies, as seen in GRB 170817A} \citep{gold17}.

Fitting Swift-XRT data to luminosity profiles appropriate for orthogonal rotators \eqref{eq:spiinlum} and precessing, oblique rotators \eqref{eq:genspinlum}, we find that the mean-square residuals can be substantially reduced (up to $\approx 220 \%$ for both GRBs 080602 and 090510), and that, despite having more free parameters, the fits are preferred, information-theoretically speaking, because the AIC values are smaller than for the orthogonal rotator fits \citep{akaike}; {see Tab. \ref{tab:fits}}. Moreover, the parameters that are pulled out from the fits are mutually consistent, and match expectations of magnetar birth; for example, the ellipticities inducing precession match well with the values expected from magnetic deformations \citep{mast15}. Although not conclusive, this provides evidence for early-time precession in millisecond magnetars born out of either supernovae (for long GRBs) or merger events (for short GRBs), and generally strengthens the millisecond magnetar proposal as an explanation for extended emission, at least in some cases.

{Indeed, it is important to note that millisecond magnetars are not the only viable explanation for X-ray afterglow curves \citep{berg14}. In fact, the standard interpretation involves the GRB jet interacting with the surrounding interstellar medium (culminating in a fireball), which produces multi-band emission \citep{mez93}. Nevertheless, plateau phases \citep{rowl13,stratta18} and/or steep falloffs at late ($t \gtrsim 10^{3}$ s) times \citep{lasky16,sarin20} are difficult to explain with a pure fireball model \citep{sarin19}, and provide motivation to study the millisecond magnetar engine for extended emissions in the X-ray band.} 

{In this work, we have focussed on the simple case of a precessing dipole. A more thorough analysis would include additional terms in \eqref{eq:spin-downrot} from neutrino-driven mass losses \citep{thom04}, gravitational radiation \citep{lasky16}, multipolar magnetic fields \citep{mast15}, and possible fall-back accretion \citep{mel14}, all of which would modify the braking index $n$ of the neutron star and help explain the inferred values $3 \lesssim n \lesssim 5$ for various GRB remnants \citep{xiao19,lu19}. In a plasma-filled magnetosphere, the braking index reads $n \approx  3 + 2 \sin^2\alpha \cos^2\alpha \left(1 + \sin^2 \alpha \right)^{-2}$ \citep{arz15}, which is generally greater than $3$ though fluctuates as $\alpha$ wobbles in a precessing model. For the fits obtained herein, this model gives $3.02 \leq n(t) \leq 3.24$ for GRB 086002 and $3.03 \leq n (t)\leq 3.23$ for GRB 090510. It would certainly be worthwhile to extend this study by including more general braking physics in future.}





\section*{Acknowledgements}
This work was supported by the Alexander von Humboldt Foundation and by the DFG research grant 413873357. {We thank the anonymous referee for their helpful feedback, which considerably improved the quality of the manuscript.}





\end{document}